# Breakdown products of gaseous polycyclic aromatic hydrocarbons investigated with infrared ion spectroscopy

A. Petrignani[*]

Leiden Observatory, Leiden University, Niels Bohrweg 2, NL-2333 CA, Leiden, The Netherlands

Radboud University, Institute for Molecules and Materials, FELIX Laboratory, Toernooiveld 7c, NL-6525 ED Nijmegen, The Netherlands

a.petrignani@science.ru.nl

M. Vala & J.R. Eyler

Department of Chemistry and Center for Chemical Physics, University of Florida, Gainesville, FL 32611-7200, USA

A.G.G.M. Tielens

Leiden Observatory, Leiden University, Niels Bohrweg 2, NL-2333 CA, Leiden, The Netherlands

G. Berden, A.F.G. van der Meer, B. Redlich, & J. Oomens

Radboud University, Institute for Molecules and Materials, FELIX Laboratory, Toernooiveld 7c, NL-6525 ED Nijmegen, The Netherlands

---

[*] Current affiliation: University of Amsterdam, Science Park 904, 1098 XH Amsterdam, The Netherlands, a.petrignani@uva.nl

# ABSTRACT

We report on a common fragment ion formed during the electron-ionization-induced fragmentation of three different three-ring polycyclic aromatic hydrocarbons (PAHs), fluorene ($C_{13}H_{10}$), 9,10-dihydrophenanthrene ($C_{14}H_{12}$), and 9,10-dihydroanthracene ($C_{14}H_{12}$). The infrared spectra of the mass-isolated product ions with m/z = 165 were obtained in a Fourier transform ion cyclotron resonance mass spectrometer whose cell was placed inside the optical cavity of an infrared free-electron laser, thus providing the high photon fluence required for efficient infrared multiple-photon dissociation. The infrared spectra of the *m/z* = 165 species generated from the three different precursors were found to be similar, suggesting the formation of a single $C_{13}H_9^+$ isomer. Theoretical calculations using density functional theory (DFT) revealed the fragment's identity as the closed-shell fluorenyl cation. Decomposition pathways from each parent precursor to the fluorenyl ion are proposed on the basis of DFT calculations. The identification of a single fragmentation product from three different PAHs supports the notion of the existence of common decomposition pathways of PAHs in general and can aid in understanding the fragmentation chemistry of astronomical PAH species.

## 1. INTRODUCTION

Gaseous polycyclic aromatic hydrocarbons (PAHs) have been studied extensively because of their role in combustion processes and infrared emission from interstellar sources (Tielens 2008). The PAHs are thought to be cycled through many different interstellar environments during their lifetimes (Allamandola et al. 1999). Most likely formed in stars on the asymptotic giant branch, they may be exposed to relatively cool photon sources, but also come into the vicinity of planetary nebulae and can be harshly irradiated by emissions from extremely hot stellar cores. In such harsh interstellar environments with high fluxes of visible/ultraviolet photons or electrons, many PAHs are expected to be ionized, isomerized, and/or fragmented. While it is understood that only the most stable PAHs can be expected to survive over the long term in these environments, such irradiation may lead to fragmentation and formation of different (and possibly more stable) species, even $C_{60}$.

Conclusive evidence has been presented that the stable carbon-cage molecule, C60, exists in a variety of stellar and interstellar environments, including planetary nebulae (Cami et al. 2010; García-Hernández et al. 2010, 2011), young stellar objects (Roberts et al. 2012), and the interstellar medium (Sellgren et al. 2010). The mechanism of its formation in these regions, however, has been a subject of intense debate. In those astrophysical sources where $C_{60}$ has been found, PAH emission is also present. As we observe closer to the star in the nebula NGC 7023, the abundance of $C_{60}$ increases rapidly while the amount of PAHs decreases (Berné & Tielens 2012). Recent studies have proposed that $C_{60}$ may be formed in a "top-down" fashion by the photochemical processing of large PAHs with sequential losses of hydrogen and carbon (Berné & Tielens 2012; Castellanos et al. 2014; Zhen et al. 2014a; Berné et al. 2015). To form fullerene from PAHs requires the conversion of some of the carbon hexagons in the planar PAH framework to five-membered rings. Such a conversion to C60 has been demonstrated convincingly in graphene via transmission electron microscopy (Chuvilin et al. 2010). Mass spectrometric analysis of the laser photolysis of large PAHs such as $C_{42}H_{18}^+$, $C_{60}H_{22}^+$, and $C_{66}H_{26}^+$ has provided evidence that production of fullerenes from certain PAHs is also possible (Zhen et al. 2014a, 2014b; Berné et al. 2015).

Over the years, a number of experimental studies have been dedicated to determining the fragmentation pattern of highly excited PAHs (Pachuta et al. 1988, Jochims et al. 1994; Allain et al. 1996; Boissel et al. 1997, 2000). Others focused on determining the initial steps in the fragmentation and the kinetic parameters involved (Dunbar 1992; Lifshitz 1997; Van- Oanh et al. 2006a, 2006b). More recently, attention has shifted toward the final fragmentation steps (Ekern et al. 1997, 1998; Joblin et al. 1997; Ling et al. 1997; Schroeter et al. 1999). Except for a few studies (Zhen et al. 2014a, 2014b; Berné et al. 2015; Bouwman et al. 2016), however, very little attention has been devoted to characterizing the intermediate fragmentation products. In this paper, we report the infrared spectra of one fragmentation product formed from the electron-ionization-induced decomposition of three PAHs. This product may be a representative of a class of intermediates necessary to the eventual formation of carbon bowls, cages, and fullerenes from larger PAHs.

In recent years, our knowledge of the vapor phase structure and reactivity of PAH ions has been significantly expanded by developments in the field of ion spectroscopy. The coupling of Fourier transform ion cyclotron resonance (FTICR) mass spectrometers and free-electron lasers (FELs) with tunable output in the infrared (such as FELIX, the FEL for Infrared eXperiments) (Oepts et al. 1995; Valle et al. 2005) has enabled the recording of the vibrational spectra and thus the determination of the structures of numerous gaseous ions, including many PAHs of astronomical relevance (Oomens et al. 2003). But, despite these valuable advances, the low absorption cross-sections and/or high dissociation thresholds of especially small to intermediate-sized ionic PAH fragments have often hindered their investigation.

To overcome these limitations, an apparatus has been developed in which the ion-trapping cell of an FTICR mass spectrometer (FTICR MS) is placed within the optical cavity of the FEL, thus offering the much higher photon fluxes available inside the laser cavity (Militsyn et al. 2003). This apparatus, referred to here as the FELICE-FTICR MS (A. Petrignani et al., *in preparation*), allows the spectral identification of fragmented and mass-selected PAH ions and the investigation of their decomposition/fragmentation pathways. We present here our study of the electron-ionization-induced fragmentation of three different PAHs—fluorene (FL), 9,10-dihydrophenanthrene (DHP), and 9,10-dihydroanthracene (DHA)—that were each found to produce a major fragment ion at *m/z* = 165 ($C_{13}H_9^+$). Here we show that this fragment cation exhibits the same infrared multiple-photon dissociation (IRMPD) spectrum irrespective of the precursor. These results are augmented by density functional theory (DFT) calculations, which point out the most probable pathways leading to this product.

# 2. METHODS

## *2.1. FELICE- FTICR / IRMPD Experiments*

The IRMPD spectra were recorded using an FTICR MS in which the ion-trapping cell was located inside the optical cavity of the free-electron laser FELICE. The experimental setup is shown in Figure 1. A full description of the apparatus (A. Petrignani et al., *in preparation*), a generic description of the application of FELs and ion-trapping mass spectrometers to obtain IR spectra of mass-selected gaseous ions (Valle et al. 2005), and details of the free-electron laser FELICE (Militsyn et al. 2003; Bakker et al. 2010, 2011) have been reported elsewhere. A brief description is given here. The FELICE cavity consists of two regions. The first, located one floor below the mass spectrometer, contains the FEL undulator with its energetic electron beam, while the second contains the guiding optics and FTICR intracavity storage cells of the mass spectrometer. The ions are produced in the ion-source region outside the laser cavity (Figure 1, bottom right). Placed in a glass container connected to the vacuum system, the PAH samples (Sigma-Aldrich) exhibited vapor pressures sufficient to allow them to sublime into the electron-ionization source chamber, where typical pressures were $\sim 10^{-7}$ mbar. Ionization, effected with 40 eV electrons, was followed by extraction of the parent and fragment ions into a quadrupole mass spectrometer (QMS) used as an RF-only ion guide. The ions were subsequently stored in a rectilinear quadrupole trap (RLT) and collisionally cooled with argon gas at room temperature and an ambient pressure of $\sim 10^{-5}$ mbar. For each parent sample, the ion optics were optimized to transmit and store the $m/z = 165$ ion fragments. For DHP, the $m/z = 179$ ion fragment was also investigated and the ion optics optimized accordingly. After exiting the RLT, the ion packet (predomi- nantly $m/z = 165$ or $m/z = 179$ ions) entered the FELICE cavity at the Jumbo quadrupole deflector (Jumbo QD), was deflected 90° into the large guiding quadrupole (LQ) along the cavity axis, and was stored in ICR cell 4, where overlap between the ion cloud and the infrared laser beam was found to be maximal. The laser beam traverses the optical cavity and is reflected by cavity mirrors. The diameter of the laser beam is largest in cell 4, decreases uniformly through cells 3 and 2, and is focused in cell 1. Voltages applied to the entrance and exit electrodes of the storage cell provide axial confinement while the 7 T magnet imposes radial confinement and high mass selectivity. Immediately after storage, and before irradiation with the IR light, unwanted masses are ejected via a stored- waveform inverse Fourier transform pulse (Comisarow & Marshall 1974; Guan & Marshall 1996), thereby isolating the $m/z = 165$ (or $m/z = 179$) ions.

The infrared spectra of the $m/z = 165$ species were investigated using IRMPD spectroscopy. The isolated ions were irradiated for 7–9 s with multiple pulses from FELICE. These pulses are $6\,\mu$s long, high-energy (0.5–5 J) macro- pulses, each consisting of picosecond-long (0.1–1 mJ) micro-pulses separated by 1 ns, with a tuning range of 5.5–17 $\mu$m at a repetition rate of 10 Hz. A small (0.5 mm) opening at the center of the end mirror of the FELICE cavity (Figure 1, top right) permits a sample of the FELICE beam to be monitored for infrared power and wavelength calibration. The sampled fraction varies according to the diameter of the FELICE beam, which changes with wavelength, and is determined using a Rayleigh length of 0.08m. When the IR wavelength is in resonance with an IR absorption band of the ion, incoherent multiple-photon absorption mediated by intramolecular vibrational redistribution can take place. The very high intensities provided by FELs, and the even higher intensities provided by FELICE, allow the ion internal energies to reach levels above the dissociation threshold, leading to fragmentation. The IR spectrum is recorded by monitoring the total fragment yield as a function of the frequency of the FELICE laser. The fragment yield is determined as the ratio of fragment ions to fragment plus parent ions and is subsequently normalized to the photon flux in cell 4, which is proportional to the power times the wavelength. The diameter of the FELICE beam will also affect the overlap between the laser beam and the ion cloud. The exact changes in overlap are unknown and not accounted for, though most probably the overlap will improve with increasing diameter. This may enhance the IRMPD efficiency and lead to more intense bands toward lower frequencies.

*2.2. Density functional theory calculations*

The equilibrium geometries and associated harmonic vibra- tional frequencies for the PAHs were calculated using DFT with the Gaussian 03 program package (Frisch et al. 2003). Becke's three-parameter hybrid functional and the non-local correction functional of Lee, Yang, and Parr (B3LYP) was used with a 6-31G(d,p) basis set (Becke 1993). The vibrational frequencies were scaled uniformly by a factor of 0.97 to correct for the effects of anharmonicity and deficiencies in the functional/basis set. This level of theory has been demon- strated to be reliable in predicting mode frequencies and relative intensities of the infrared absorption of medium-sized PAHs (Langhof 1996; Bauschlicher 1998). The energies given in the potential energy diagrams shown in Figures 3–5 are all corrected for zero-point energy but not scaled. All transition state (TS) structures have been determined to yield one imaginary frequency whose vibrational motion is appropriate for the product being formed. The QST3 procedure was used to locate difficult-to-find TSs (Frisch et al. 2003).

## 3. RESULTS AND DISCUSSION

*3.1. Fluorene*

*3.1.1. Experimental spectrum of the m/z=165 fragment ion from FL*

Electron ionization of fluorene leads to the formation of the parent radical cation as well as multiple fragment ion peaks at lower $m/z$. In order of decreasing peak intensity, we observe the ion resulting from H-atom loss, $m/z = 165$, the fluorene radical cation, $m/z = 166$, and the product ion with three H-atom losses at $m/z = 163$. Although other masses have been observed previously in the electron ionization of fluorene (Linstrom & Mallard 2015), they were insignificant in the present experiments due to the optimization of the ion optics to the higher masses. The singly dehydrogenated $m/z = 165$ product ion, $C_{13}H_9^+$, was mass-isolated in the FTICR MS and its IRMPD spectrum was recorded to investigate its structure. IRMPD-induced fragment ions were predominantly the H-loss channels, in order of decreasing intensity $m/z = 164, 163, 161$, followed by 162 ($C_{13}H_n^+$ with n = 5–8). Smaller contributions were observed at $m/z = 85, 87$, and 89 ($C_7H_{2n+1}^+$ with n = 0–2) and $m/z = 61$–63 ($C_5H_n^+$ with n = 1–3). Traces of lighter fragment ions were observed at, e.g., $m/z = 39$ ($C_3H_3^+$).

Previous experiments on the fluorene radical cation ($C_{13}H_{10}^+$, $m/z$ = 166) using the infrared output of a FEL, but without the intracavity setup, showed similar H-loss channels, but not the C-loss channels because the photon densities were not high enough to reach the high internal energies required (Valle et al. 2005). Replacing the FTICR setup with a 3D-quadrupole ion trap increased the photon density as a result of tighter focusing and led to the observation of $C_2H_m$-loss channels; H-loss channels were not resolved in this study (Oomens et al. 2001). Unlike these previous studies, the present apparatus has both sufficient mass-resolving power to observe the single H-loss channels and sufficient laser power to reach the high internal energies necessary to observe the C-loss channels.

The IRMPD spectrum of the $m/z$ = 165 product ion, shown in Figure 2, shows a clear double band structure around 700–800 cm$^{-1}$. In addition, it exhibits roughly four over-lapping bands between 1000–1100, 1100–1300, 1300–1500, and 1500–1700 cm$^{-1}$. The observed line widths are broad (60–100 cm$^{-1}$), which can be attributed to the very high photon densities provided by FELICE leading to very high internal excitation (A. Petrignani et al., *in preparation*).

### *3.1.2. Theoretical spectrum of the fluorenyl ion and its formation*

The calculated spectrum of the fluorenyl ion [FL–H(9)]$^+$ is shown in Figure 2. The observed spectrum of the $m/z$ = 165 product is in good agreement with the calculated spectrum, showing that the $m/z$ = 165 fragment ion is the fluorenyl cation. The large differences in intensity between experiment and theory, specifically toward lower frequencies, are partly due to the changing overlap of the ion cloud with the FELICE beam (see Section 2.1). Also, deviations in line intensities and broadening are expected as the calculated spectrum is based on single-photon absorption whereas the experiment entails the nonlinear absorption of multiple photons (Bagratashvili et al. 1985; Rijs & Oomens 2015).

Using FTICR mass spectrometry, Dibben and coworkers studied the visible/UV photodecomposition of the fluorene cation $C_{13}H_{10}^+$ and determined that one to five hydrogen atoms were lost sequentially (Dibben et al. 2001). Subsequent theoretical work (Szczepanski et al. 2001) showed that loss of an H from the sp$^3$-hybridized carbon requires considerably less energy than loss of any of the sp$^2$-hydrogen atoms (2.64 eV versus 4.0–4.9 eV (B3LYP/6-31G (d, p) level)). No barrier to dissociation of the C–H(9) bond was found. Van-Oanh and coworkers investigated the photofragmentation of the fluorene cation using an innovative approach involving a laser-irradiated supersonic molecular beam interrogated by time-of-flight mass spectrometry to determine the evolution of the rate constant for the dominant H-loss channel (Van-Oanh et al. 2006a, 2006b). All these studies lead us to conclude that the H atom lost in the present decomposition process is one of the sp3-carbon hydrogen atoms, thus producing the fluorenyl cation.

The IR photodissociation fragment of the fluorenyl ion observed at $m/z = 87$ could be the loss of acetylene, followed by the loss of $C_4H_4$, a route suggested previously (Szczepanski et al. 2001). Similarly, the additional fragment traces at $m/z = 89$ could be reached by loss of $C_4H_2$, followed by loss of $C_2H_2$. The latter route can result in the $m/z = 63$ fragment ion via another acetylene loss, observed here in small but significant intensities.

### *3.2. Dihydroanthracene*

#### *3.2.1. Experimental spectrum of the m/z=165 fragment ion from DHA*

The electron ionization of DHA also leads to the observation of multiple mass peaks, with one of the major peaks corresponding to loss of [C+3H], i.e., $m/z = 165$. Similar to the experiments with FL, all settings were optimized to produce, guide, store, and mass-isolate the $m/z = 165$ ionization product. The fragment masses produced by IRMPD were dominantly ions at $m/z = 164$ and 163, a small contribution at 161, followed by smaller contributions at $m/z = 85, 87$, and 89 ($C_7H_{2n+1}^+$ with n = 0–2) and $m/z = 61$–63 ($C_5H_n^+$ with n = 1–3). Traces of smaller masses were found at, e.g., $m/z = 39$ ($C_3H_3^+$). In fact, the same fragment ions were produced from FL$^+$, but with different relative intensities. For the $m/z = 165$ ion from DHA, only two H losses were observed, and the production of the $m/z = 39$ fragment was more clearly observable.

The IRMPD spectrum of the $m/z = 165$ ion from DHA is shown in Figure 2. The spectrum is remarkably similar to the FL $m/z = 165$ ion spectrum. It also shows a clear double band structure around 700–800 cm$^{-1}$ and exhibits similar but slightly different features of overlapping bands between 1000 and 1700 cm$^{-1}$. The observed differences are mostly in the signal-to-noise ratio, which can be attributed to differing ion numbers, laser power, and overlap between the laser and the ion cloud.

### 3.2.2. Decomposition pathways to the m/z=165 fragment ion

The DHA parent leading to the $m/z = 165$ fragment first undergoes ionization to DHA$^+$ and then dissociation. Three possible decomposition pathways for DHA$^+$ ($m/z = 180$) can be envisaged. The reaction enthalpies (B3LYP/6-31G (d, p), corrected for zero-point energy) for each step in the three schemes are given below. In pathway I, a neutral CH$_2$ fragment is ejected, forming the fluorene cation, which then loses one H to give the sought-after fluorenyl ion [FL–H(9)]$^+$:

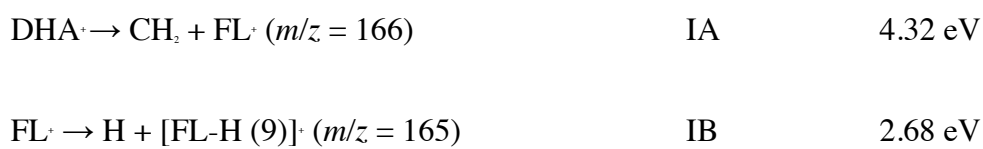

| | | |
|---|---|---|
| DHA$^+$ → CH$_2$ + FL$^+$ ($m/z = 166$) | IA | 4.32 eV |
| FL$^+$ → H + [FL-H (9)]$^+$ ($m/z = 165$) | IB | 2.68 eV |

In pathway II, a hydrogen atom is removed from position 9 (or 10) of DHA$^+$, leaving the 10 (or 9)-hydroanthracene ion. Removal of the CH$_2$ group finally gives [FL–H(9)]$^+$:

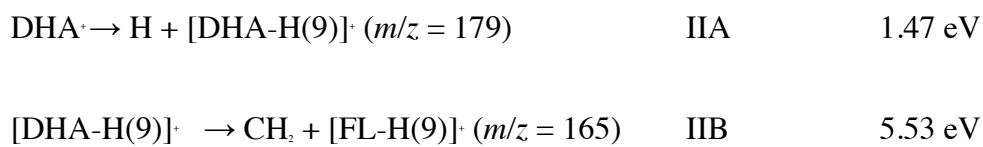

| | | |
|---|---|---|
| DHA$^+$ → H + [DHA-H(9)]$^+$ ($m/z = 179$) | IIA | 1.47 eV |
| [DHA-H(9)]$^+$ → CH$_2$ + [FL-H(9)]$^+$ ($m/z = 165$) | IIB | 5.53 eV |

In pathway III, the sequential loss of H, CH, and a further H also produces the desired $m/z = 165$ fragment ion:

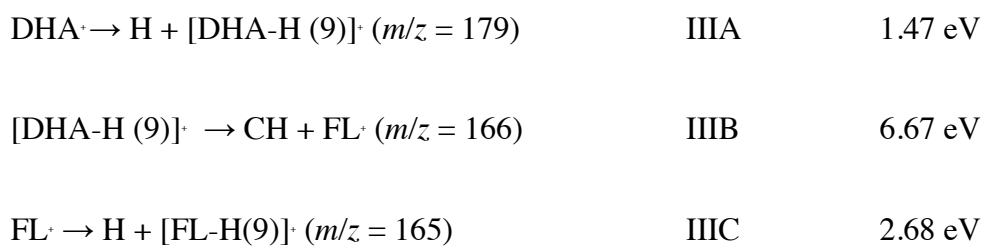

| | | |
|---|---|---|
| DHA$^+$ → H + [DHA-H (9)]$^+$ ($m/z = 179$) | IIIA | 1.47 eV |
| [DHA-H (9)]$^+$ → CH + FL$^+$ ($m/z = 166$) | IIIB | 6.67 eV |
| FL$^+$ → H + [FL-H(9)]$^+$ ($m/z = 165$) | IIIC | 2.68 eV |

Several steps in the above pathways are identical (e.g., IB and IIIC, and IIA and IIIA). The total energies for schemes I and II are identical (7.00 eV). The energy for III (10.82 eV) is larger by 3.82 eV, which represents the energy required to break one of the CH bonds in the CH2 moiety, a cleavage that does not occur in either I or II. While these results seem to favor scheme I or II, calculations do not allow a further narrowing. In scheme II ions are expected at *m/z* = 179 and 165, but not 166. In schemes I and III, fragment ions are also expected at *m/z* = 166. Our mass spectrometric results show that fragment ions from the electron ionization of DHA were observed at *m/z* = 179, 178, and 165, but not at *m/z* = 166. Although a small amount of the *m/z* = 166 product is known to form upon electron ionization (Linstrom & Mallard 2015), this includes the ionization of a 13C isotopologue, which we can eject from the storage cell before irradiation. We therefore preliminarily conclude that scheme II is the most probable pathway for the electron-ionization decomposition of DHA+.

### *3.2.3. Potential energy surfaces for fluorenyl ion formation*

The pathway of scheme II to the fluorenyl ion from DHA+ involves two steps. In IIA, one H is removed from position 9 to give the 10H-anthracene ion, followed by the extraction of $CH_2$ from position 10, which then closes to form the central five-membered ring. To determine whether either of these steps involves an activation barrier, scans of potential energy surfaces (PESs) were computed for each. For the ejection of an H atom from position 9, all degrees of freedom were optimized except the incremented C–H bond length. The dissociation energy for the H loss (from structure A (DHA+) to B (10H-anthracene+) is 1.51 eV, a value close to the reaction energy of 1.47 eV, indicating a zero or small barrier to the extraction.

Figure 3 shows the relative energies and molecular structures of the species produced from B. At each step, all degrees of freedom were optimized except the incremented bonds. Increasing one of the C–$CH_2$ bonds to 1.64 Å (with an energy input of 2.84 eV) results in C, a TS whose two aromatic rings are now non-coplanar. One imaginary frequency was found whose motion indicates five-member ring closure. Closure of this ring results in a stabilization of 0.39 eV and leads to D, a stable fluorenyl species with a $CH_2$ attached to carbon C(b); the angle C(c)–C(b)–C(10) is 65.6°. Two different types of calculation were performed on this structure. First, a non-relaxed computation of the ejection of the $CH_2$ group required 4.03 eV and yielded fluorenyl and $CH_2$. Second, a relaxed scan produced the TS G, in which the angle C(c)–C(b)–C(10) has increased to 85.2° (with an input of 0.48 eV). Further increasing the C(b)–C(10) distance results in a shift of the methylene group and forms H, which now has a three-membered ring attached to a six-membered ring across the C(b)–C(4) carbons, with a resultant 1.19 eV stabilization. With a further input of 0.5 eV energy, the C(b)–C(4) bond opens and a seven-membered ring forms, giving I. From intermediate D, a three-membered ring may form with a slight stabilization to form E.

### *3.3. Dihydrophenanthrene*

#### *3.3.1. Electron Ionization of DHP*

The electron ionization of DHP leads to the observation of multiple fragment ions, resulting from both H- and C-loss channels. Of the H-loss channels, the $m/z = 179$ product (single H loss), dominates. In addition, both $m/z = 166$ ions (C +2H loss) and $m/z = 165$ ions (C+3H loss), as well as $m/z = 152$ ions (2C+4H loss), were observed. No fragment ions with lower masses were observed, contrary to FL and DHA, which can be attributed to the fact that the ion optics were optimized to transmit the $m/z = 165$ and 179 fragment ions. In addition to the IR spectrum of the $m/z = 165$ fragment ion, the spectrum of the $m/z = 179$ fragment ion was also recorded (see Appendix).

##### *3.3.1.1. The m/z=165 fragment ion*

The fragment masses produced by IRMPD of the $m/z = 165$ fragment ion were mainly $m/z = 164$ and 163, with a smaller peak at $m/z = 161$. Weaker peaks were observed at $m/z = 85, 87$, and 89 ($C_7H_{2n+1}^+$, n = 0–2) and $m/z = 61$–63 ($C_5H_n^+$, n = 1–3). Traces of lighter fragment ions were observed at, e.g., $m/z = 53$ ($C_4H_5^+$) and $m/z = 39$ ($C_3H_3^+$). Small differences in fragment intensities from DHA+ were also observed in the C-loss channels. No 2C+mH loss was observed.

The IRMPD spectrum of the $m/z = 165$ ion from DHP+ is shown in Figure 2. This spectrum is similar to those found for the $m/z$ 165 fragment ions from FL and DHA. The double band structure at 700–800 $cm^{-1}$ is clearly present as are the overlapping bands between 1000 and 1700 $cm^{-1}$. As before, the observed differences can likely be attributed to relatively low signal-to-noise ratios and possibly to differences in internal energies.

In a recent study using tandem mass spectrometry and imaging photoelectron–photoion coincidence (iPEPICO), West and coworkers (West et al. 2014) investigated the $m/z = 165$ decomposition product from DHP. These authors demonstrated that the $m/z = 165$ mass spectrum from DHP$^+$ due to metastable ion–collision-induced dissociation (MI-CID) was identical to the $m/z = 165$ hydrogen-loss mass spectrum from FL$^+$, and concluded that the decomposition product from both ions was the fluorenyl ion, in agreement with the present spectroscopic results.

*3.3.2. Theory of the m/z = 165 ion and its formation from DHP$^+$*

The formation of the $m/z = 165$ fragment from the DHP$^+$ parent requires the loss of 15 mass units, which could involve a single $CH_3$ moiety. Numerous calculations were performed to determine whether a viable pathway could be found for the ejection of a single $CH_3$ group (e.g., formed by a departing $CH_2$ group attracting one H to form an ejected $CH_3$ group). However, none was found.

The pathways described above for DHA$^+$ (Equations (IA)– (IIIC)) can also be used to describe the decomposition pathways possible for DHP$^+$, by substituting DHP$^+$ for DHA$^+$. The reaction enthalpies (in eV) for DHP$^+$ are: IA, 4.80; IB, 2.68; IIA, 2.30; IIB, 5.18; IIIA, 2.30; IIIB, 6.32; IIIC, 2.68.

Three observations/deductions are useful in choosing the most probable pathway. First, the major decomposition product observed from DHP$^+$ is the $m/z = 165$ species, identified here as the fluorenyl ion, as well as a secondary product at m/z = 179. Second, the measured decomposition of the m/z = 179 fragment yields $m/z = 178$ and 152 fragments, but no $m/z = 165$ and 166 species. Third, since the major $m/z = 165$ fragment was observed experimentally upon decomposition of DHP$^+$, it must originate from the $m/z = 180$ parent molecule.

Scheme I provides a possible route to both the $m/z = 166$ and 165 products, while scheme IIA (or equivalently, IIIA) can account for the $m/z = 179$ product. Clearly, schemes IIB and IIIB are not operative since the $m/z = 165$ and 166 products do not originate from the $m/z = 179$ precursor.

The computed PES for scheme I is shown in Figure 4(a). Elimination of a $CH_2$ group from position 10 requires 7.28 eV (A to TS B), where the C(a)–C(10) distance in the TS B is 3.70 Å and the C(9)–C(10) distance is 3.58 Å. Closure of the five-membered ring stabilizes the $m/z = 166$ fragment (C) by 2.48 eV. Finally, an input of 2.68 eV is required to eject one of the sp$^3$-hybridized hydrogen atoms, giving the final fluorenyl ion D plus a free CH$_2$ group and H atom.

The potential energy diagram for scheme II is shown in Figure 4(b), where the 10H-phenanthrene ion E has been formed from the parent DHP$^+$ (A). The removal of one H is computed to require 2.31 eV, a value identical to the thermodynamic value (IIa), signifying zero activation barrier to this decomposition.

Forming the $m/z = 166$ and 165 products from E requires the removal of CH and CH$_2$ groups. As Figure 4(b) shows, increasing the C(d)–C(9) bond (from 1.42 to 1.87 Å) requires 2.47 eV (E to TS F). With a further increase in the C(d)–C(9) bond to 1.92 Å, the intermediate G is formed, with the H atom now completely shifted from C(10) to C(9). Further increasing the C(d)–C(9) distance to 2.3 Å leads to the TS H, which leads spontaneously to the formation of a bond between C(10) and C(d), i.e., closure of the five-membered ring, and the stable structure I, the 10-methylenefluorenyl cation. With an input of 5.70 eV, the CH$_2$ group is ejected, leaving the free fluorenyl cation D and CH$_2$.

Instead of breaking the C(d)–C(9) bond, the C(a)–C(10) bond may also rupture. This requires 2.45 eV (E to TS H), and thence to intermediate I, and finally to D. To reach D from I requires removal of the methylene group. This calculation involving the removal of the methylene group (I to D) was approached in two ways. First, the C(9)–C(10) bond length was incremented (0.2 Å steps) without optimizing the remainder of the structure. This approach requires 5.70 eV to remove the CH$_2$ (Figure 4(b)). Second, when optimization of the remainder of the structure is permitted, the results are very different. As Figure 5(a) shows, and as is discussed fully in the Appendix, structures are formed containing five-, six-, and seven-membered rings fused to one another.

# 4. ASTROPHYSICAL IMPLICATIONS

Ever since the discovery of $C_{60}$ 30 years ago, many proposals on its mechanism of formation have been reported. Most have invoked formation from carbon atoms or small carbon clusters, chains, or rings, so-called bottom-up mechanisms (Kroto & McKay 1988; Heath 1992; Hunter et al. 1994; Dunk et al. 2013). But recently a top-down approach has been suggested (Chuvilin et al. 2010; Zhang et al. 2013). First seen in the aberration-corrected transmission electron microscopy of graphene (Chuvilin et al. 2010), these experiments showed directly the loss of edge carbon atoms by the electron beam, reconstruction (i.e., formation of carbon pentagons) at that site, curling of the sheet, and ultimate creation of the fullerene. Subsequent studies of large PAHs have also shown that the top-down model can produce fullerenes (Castellanos et al. 2014; Zhen et al. 2014a; Berné et al. 2015) and that these processes are possible in certain environments in space (Berné & Tielens 2012). In this model, the dehydrogenation of the PAHs by UV photoabsorption leads to graphene-like structures (loss of multiple edge hydrogen atoms). Further irradiation results in carbon loss at the zigzag edge of the graphene-like structure, which in turn leads to rearrangement and pentagon formation at its edges. Pentagon formation is, of course, essential to the eventual curling or zipping-up of the molecule. Many different organic structures can be expected along this route.

In the present work we have shown that electron-induced decomposition of three moderate-sized three-ring PAHs can lead to the same fragment ion, the fluorenyl cation. In addition to the formation of a product that contains a five-membered ring, our result shows that stable PAH species may be converted into other stable species under appropriate excitation conditions and that the resulting photo- or electron-induced products are not necessarily the thermodynamically most stable forms. It is interesting to speculate whether the photodecomposition of a wide variety of PAHs might lead to the formation of a limited number of stable intermediates. Observationally, there is good evidence for the dominance of a limited number of PAH species (so-called grand-PAHs) in the harshest environments of space (Andrews et al. 2015). The present results, though limited, provide experimental support for this fragmentation. Clearly, further experimental studies are warranted and should focus on addressing this question for the initial fragmentation steps of larger PAHs to determine the competition between fragmentation and isomerization with respect to the formation of carbon cages and fullerenes.

## 5. CONCLUSIONS

An apparatus incorporating an FTICR cell within the optical cavity of a FEL with tunable wavelength in the infrared region has been utilized to record the IRMPD spectra of the species produced in the electron-ionization decomposition of three different PAHs: fluorene, 9,10-dihydroanthracene, and 9,10-dihydrophenanthrene. Interestingly, all three yield the same m/z = 165 product ion. The experimental infrared spectra of these ions are remarkably similar, supporting the conclusion that all three $m/z$ = 165 ions possess the same molecular structure, shown here to be the fluorenyl cation.

DFT computations at the B3LYP/6-31G (d, p) level have been used to probe the possible decomposition pathways from the three PAH precursors. It is shown that the most probable decomposition pathway for DHA$^+$ involves the ejection of a hydrogen atom from position 10 of the central six-membered ring, followed by breaking of the C–CH$_2$ bonds and subsequent formation of the five-membered ring of the fluorenyl product.

The decomposition of DHP$^+$ is shown to involve several different pathways. Extraction of a CH$_2$ group from position 9 (or 10) of DHP$^+$ followed by closure to a central five-membered ring yields the fluorene cation. Subsequent ejection of a hydrogen atom leads to the fluorenyl product. On the other hand, initial ejection of the hydrogen atom from position 9 (or 10) yields the $m/z$ = 179 product that may further decompose via two channels to the 9-methylenefluorene ion. Subsequent ejection of the CH$_2$ group could lead to the fluorenyl ion, but a lower-energy rearrangement of the 9-methylenefluorene ion to two different $m/z$ = 179 isomers appears to be preferred.

The finding that three different PAHs can produce the same fragmentation product containing a five-membered ring may have important consequences for our understanding of the conversion of PAHs and fullerenes in space.

This work was supported through the Advanced ERC grant 246976 from the European Research Council, through a grant by the Dutch Science Agency, NWO, as part of the Dutch Astrochemistry Network, and through the Spinoza prize from NWO. The work was also part of the research program of the Stichting voor Fundamenteel Onderzoek der Materie (FOM), which is financially supported by NWO. Part of the FELIX facility was financed through the NWO BIG program. Travel support was provided to J.R.E. from the National Science Foundation under Grant OISE-0730072. We greatly appreciate the skillful assistance of the FELIX facility staff.

**APPENDIX:** *The [DHP – H]$^+$ ion at m/z = 179*

In this appendix we discuss (1) the rationale behind the nonappearance of the *m/z* = 165 and 166 fragments in the decomposition of the *m/z* = 179 fragment ion [DHP–H(9)]$^+$ E, (2) the possible formation pathways to various isomers of E, and (3) speculation over the importance of these types of isomers in the degradation of other PAHs.

When full optimization is permitted in the removal of the methylene group in I (except for the incremented CH–CH2 bond distance) (Figure 4(b)), structure J is obtained (Figure 5(a)). With an energy difference of only 0.07 eV, I and J are most probably in equilibrium. As Figure 5(a) shows, with an input of 1.35 eV, the CH2 shifts to the nearby C(a), the carbon atom common to the both five- and six-membered rings in TS K. Upon formation of a three-membered ring with C(1), K is stabilized and forms L. Two pathways are now possible. Breaking the C(a)–C(1) bond leads through TS N to P, the 1-hydrocycloheptatriene-2,3-indenyl ion. The other possibility, opening the C(a)–C(9) bond of the three-membered ring, forms TS M. From there, a new three-membered ring may form with a bond between C(9) and C(2), giving the stable structure O. Although not pictured, this species may undergo a similar opening of the three-membered ring to produce the 2- hydrocycloheptatriene-2,3-indenyl ion.

The total barriers in this latter approach are significantly smaller than found using the first approach, as expected for a relaxed versus a non-relaxed scan. Indeed, this approach does not lead to the removal of the CH$_2$ group, but simply yields a number of isomers of the parent 10H-phenanthrene cation, namely, (a,1)-cyclomethylenefluorenyl L and 1-hydrocyclo- heptatriene-2,3-indenyl ion P. The question now is, is there any evidence for these isomers?

The IRMPD spectrum of the *m/z* = 179 ion is shown in Figure 5(b). Fragments from the *m/z* = 179 ion include (in order of decreasing intensity) *m/z* = 178, 152, 150, 151, 153 ions with the H-loss channel dominant, followed by loss of 2C+5H. No loss of C+2H or C+3H was observed, which would have created *m/z* = 165 or 166 products. The IRMPD spectrum shows a strong resonance at ~760 cm$^{-1}$, a broad band of low intensity at 1100–1300 cm$^{-1}$, a stronger broad band at 1300–1400 cm$^{-1}$, and a very broad structure at 1400–1700 cm$^{-1}$. Figure 5(b) also displays the computed spectra of the three isomer ions: 10-hydrophenanthrene E, (a,1)-cyclomethylenefluorenyl L, and the 1-hydrocyclohepta- triene-2,3-indenyl P, and their sum. Although the observed intensity between 1500 and 1600 cm$^{-1}$ may hint at the presence of isomers (e.g., L and/or P) other than 10-hydrophenanthrene, the limited resolution of the experimental spectrum does not warrant any definitive conclusion at present.

Despite the lack of evidence for any isomers other than E, it is now clear why the CH and CH$_2$ groups are not ejected from 9-methylenefluorenyl ion isomer I (see Figure 4(b)). Breaking the C(d)–C(9) bond in 10-hydrophenanthrene ion E leads to the spontaneous 1, 2 H-shift from C(10) to C(9) (E to TS F to G). This is a consequence of the greater stability achieved when the CH group is adjacent to the six-membered ring rather than at the end of the pendant CH$_2$–CH group, and of the lower-energy pathway to the L and P isomers than the full CH$_2$ ejection. The pathway to CH2 ejection E–H–I–D requires 6.67 eV whereas the formation of the three *m/z* = 179 isomers via E–H–I–J–K requires 1.62 eV to reach L, 1.36 eV to produce P, and 1.56 eV to reach O (see Figures 4(b) and 5(a)).

An interesting feature of the 1-hydrocycloheptatriene-2,3-indenyl ion P isomer and a similar isomer formed from DHA$^+$ (I in Figure 3) is their unusual structure, consisting of five-, six-, and seven-membered rings fused to one another. Such structures are reminiscent of electron-bombarded, reconstructed graphenes and of defects in the fullerenes. In their study of the stability of graphene edges, Koskinen and coworkers predicted that reconstructed edges containing fused five- and seven-membered rings are more common than originally thought and are more stable than the common zigzag pattern (Koskinen et al. 2009). This so-called "reczag" configuration was confirmed in subsequent experiments (Chuvilin et al. 2009). The effect of defects (i.e., the inclusion of five- and seven-membered rings) on vibrational spectra of a number of large PAHs has been computed using DFT approaches (Ricca et al. 2011; Yu & Nyman 2012). Very recently Bouwman and coworkers demonstrated experimentally that the dissociative ionization of naphthalene leads to ejection of acetylene and formation of the pentalene cation, which contains two fused five-membered rings (Bouwman et al. 2016). Scuseria and coworkers have calculated the role of $sp^3$ carbon atoms and seven-membered rings on the bond rearrangements and fragmentation in high-energy processes in various fullerenes (Eckhoff & Scuseria 1993; Murry et al. 1993; Xu & Scuseria 1994). Whitesides and Frenklach have detailed the influence of five-membered carbon rings on the growth and curvature of large graphene sheets, starting from linear or compact PAHs (Whitesides & Frenklach 2010). Our theoretical proposal of a reczag-like molecular isomer P as a decomposition product of DHP should be investigated further experimentally to determine its possible importance in the photodegradation of larger PAHs.

Ekern, S. P., Marshall, A. G., Szczepanski, J., & Vala, M. 1997, ApJL, 488, L39

Ekern, S. P., Marshall, A. G., Szczepanski, J., & Vala, M. 1998, JPCA, 102, 3498

Frisch, M. J., Trucks, G. W., Schlegel, H. B., et al. 2003, Gaussian 03, Revision B.05 (Wallingford, CT: Gaussian, Inc.)

García-Hernández, D. A., Kameswara, R. N., & García-Hernández, D. L. 2011, ApJ, 729, 126

García-Hernández, D. A., Manchado, A., García-Lario, P., et al. 2010, ApJL, 724, L39

Guan, S., & Marshall, A. G. 1996, IJMSI, 157/158, 5

Heath, J. R. 1992, in ACS Symp. Series, Fullerenes: Synthesis, Properties, and Chemistry of Large Carbon Clusters, ed. G. Hammond & V. Kuck (Washington, DC: American Chemical Society), 1

Hunter, J. M., Fye, J. L., Roskamp, E. J., & Jarrold, M. F. 1994, JPhCh, 98, 1810

Joblin, C., Masselon, C., Boissel, P., et al. 1997, Rapid Commun. Mass Spectrom., 11, 1619

Jochims, H. W., Rühl, E., Baumgärtel, H., et al. 1994, ApJ, 420, 307 Koskinen, P., Malola, S., & Häkkinen, H. 2009, PhRvB, 80, 073401

Kroto, H. W., & McKay, K. 1988, Natur, 331, 328

Langhof, S. R. 1996, JPC, 100, 2819

Lifshitz, C. 1997, Int. Rev. Phys. Chem., 16, 113

Ling, Y., Martin, J. M. L., & Lifshitz, C. 1997, JPCA, 101, 219

Linstrom, P. J., & Mallard, W. G. (ed.) 2015, NIST Chemistry WebBook, NIST Standard Reference Database Number 69 (Gaithersburg, MD: National Inst. of Standards and Technology) http://webbook.nist.gov, (retrieved July 14)

Militsyn, B. L., von Helden, G., Meijer, G. J. M., & van der Meer, A. F. G. 2003, NIMPA, 507, 494
21

# Figures

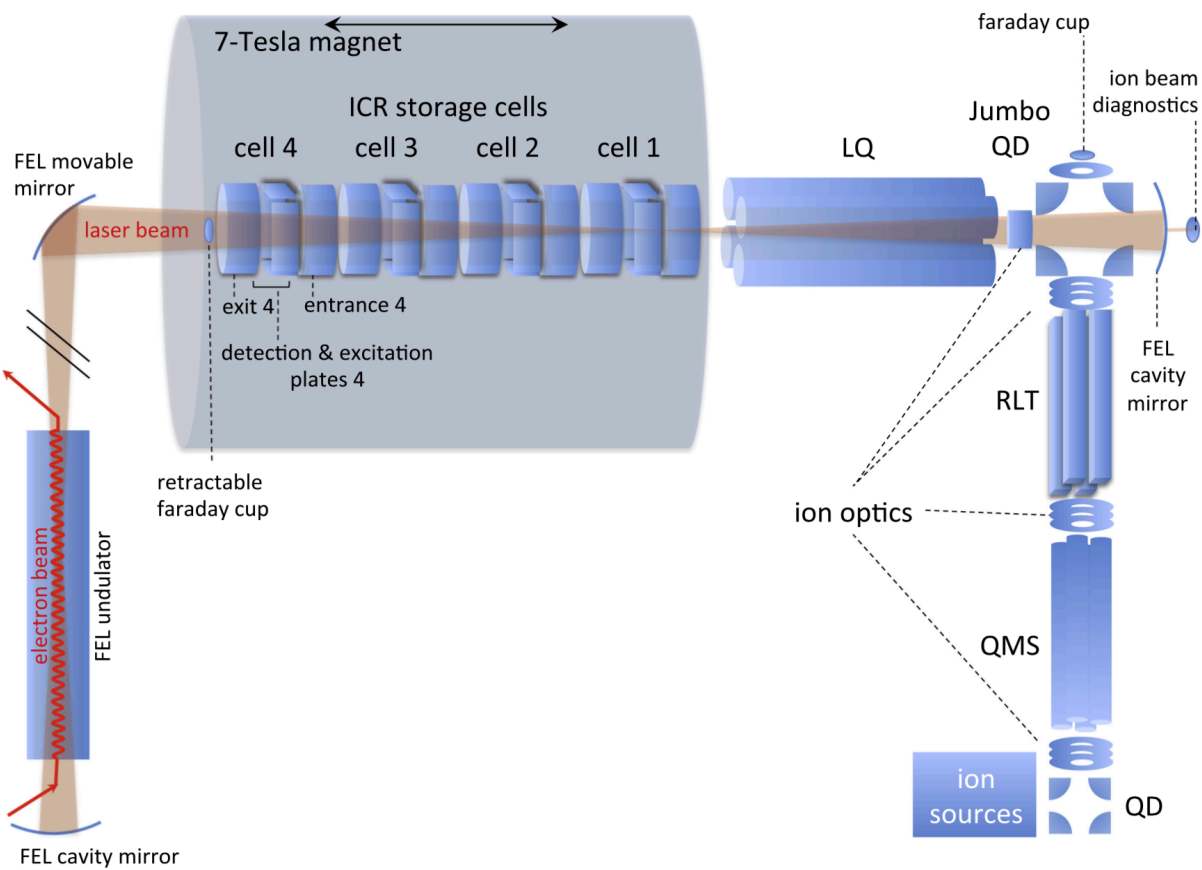

**Figure 1.** Schematic diagram of the FELICE-FTICR mass spectrometer. The PAH ions were produced on the bottom right and were guided, deflected, focused, and pulsed using various ion optics including a quadrupole deflector (QD), quadrupole mass spectrometer (QMS), rectilinear ion trap (RLT), jumbo quadrupole deflector (Jumbo QD), and a large guiding quadrupole (LQ), after which they were stored in storage cell 4 for measurements with the free-electron laser FELICE. For further details, see A. Petrignani et al. (*in preparation*).

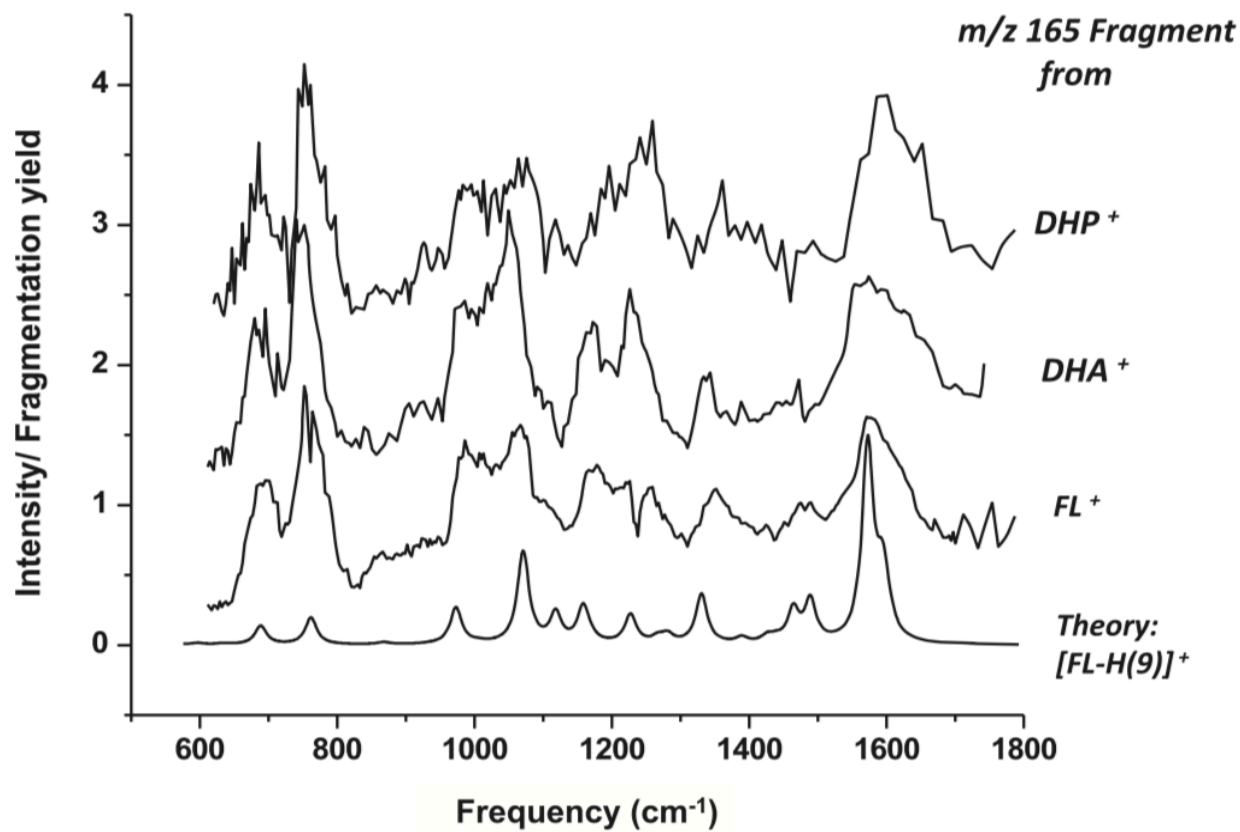

**Figure 2.** Experimental infrared multiple-photon dissociation (IRMPD) spectra of the *m/z* = 165 fragment ion obtained from electron-ionization fragmentation of the 9,10-dihydrophenanthrene ion (DHP⁺), 9,10 dihydroanthrancene ion (DHA⁺), and fluorene ion (FL⁺) and the predicted infrared absorption spectrum (B3LYP/6-31G (d, p) level) of the fluorenyl ion ([FL–H(9)]⁺). The computed spectrum (B3LYP/6-31G (d, p)) was scaled by 0.97 and a spectral bandwidth of 10 cm⁻¹ was adopted.

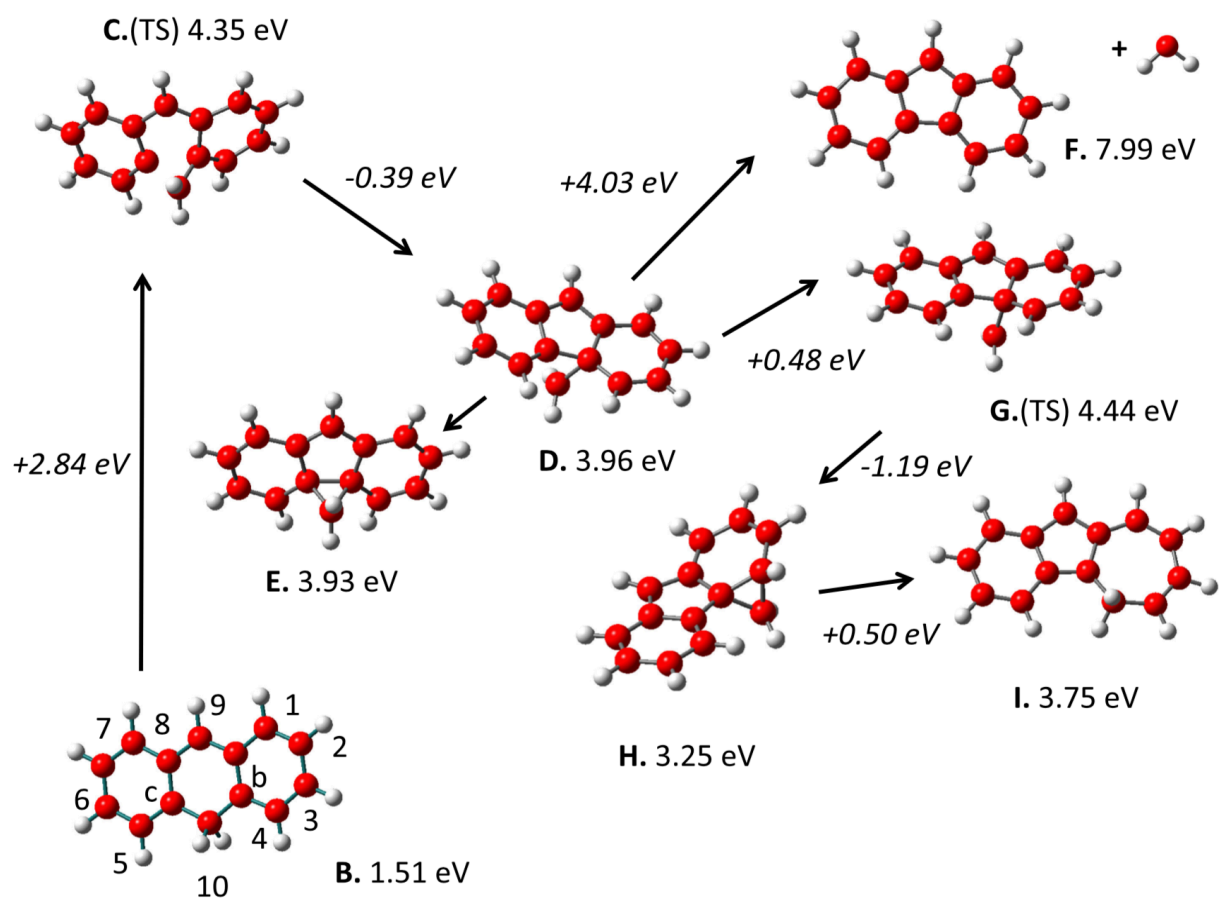

**Figure 3.** Potential energy diagram (B3LYP/6-31G (d, p) level) for the dissociation and isomerization of the 10-hydroanthracene cation ([DHA–H(9)]⁺, B. Excitation of B leads to the TS C in which a C–C bond in the central ring is broken and the two outer rings are folded in butterfly fashion. Closure of the central five-membered ring stabilizes the structure, leading to the intermediate D, the b-methylene-fluorenyl ion. A further input of 4.03 eV leads to the ejection of the $CH_2$ group from the b-position yielding F, the fluorenyl ion. Lower-energy pathways lead to two isomers, E and I. In E a three-membered ring is attached to the five-membered one. In a different pathway, a shift of methylene produces the TS G, which then forms the intermediate H; the ring may then open, yielding the 5-hydrocycloheptatriene-indenyl ion, I.

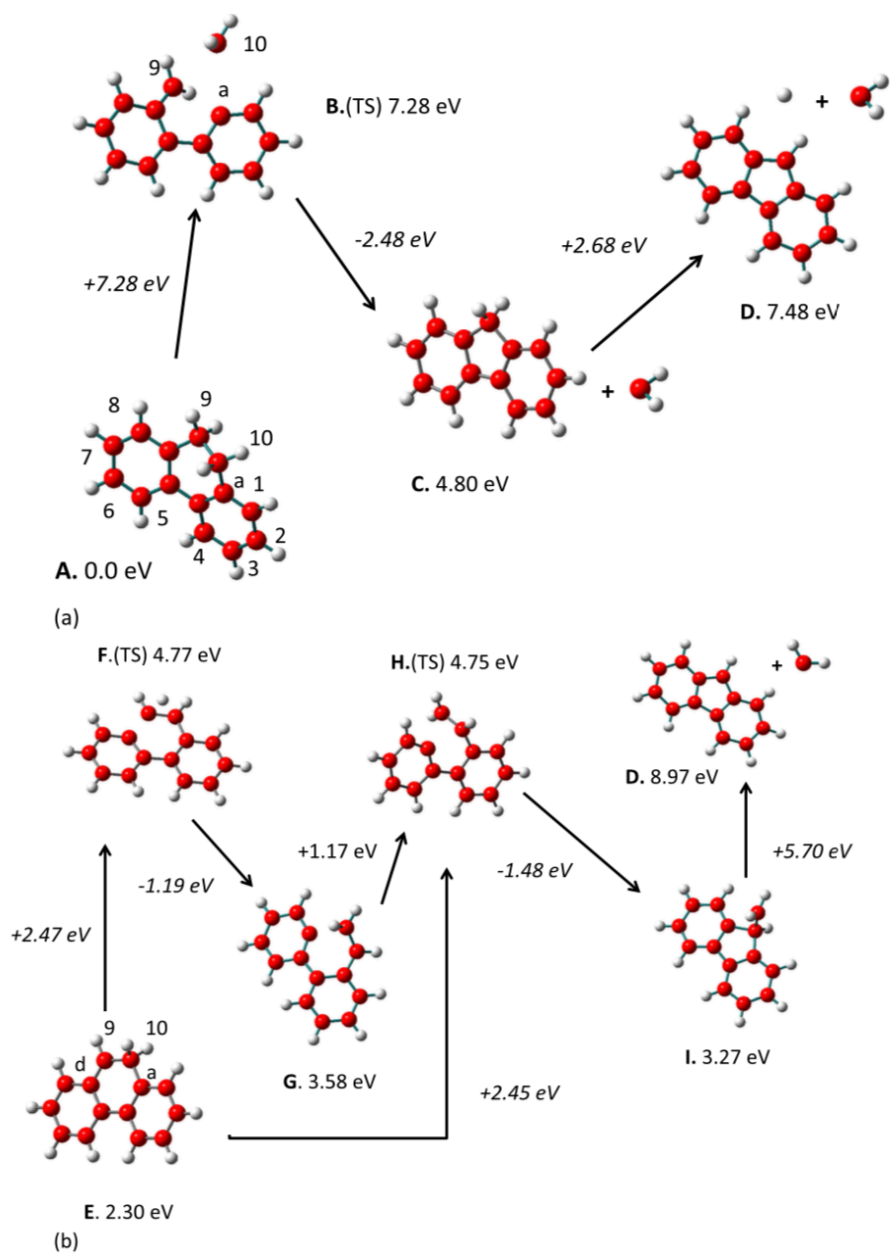

**Figure 4.** (a) Potential energy diagram (B3LYP/6-31G (d, p) level) for the decomposition of 9,10-dihydrophenanthrene, A (DHP·, $m/z$ = 180), leading to TS B, in which a $CH_2$ is ejected and which, upon closure, yields the fluorene ion, C, and $CH_2$. One H from position 9 is then ejected, giving the fluorenyl ion D plus $CH_2$ and H. (b) Potential energy diagram for the dissociation of the 10-hydrophenanthrene ion E ([DHP–H(9)]·, $m/z$ = 179), produced by the ejection of an H from position 9 of DHP· A. Two dissociation pathways are possible. Rupture of the C(d)–C(9) bond leads to the TS F in which 1, 2 H-shift of H(10) to C(9) yields the intermediate G. Twisting the C(9)$H_2$ group out-of-plane generates the TS H. In the other pathway, breaking the C(a)–C(10) bond in E leads directly to TS H. From H, close approach of the C(d) and C(10) carbon atoms yields the stable structure I, 9-methylenefluorene. Ejection of $CH_2$ from I forms the fluorenyl ion D plus $CH_2$.

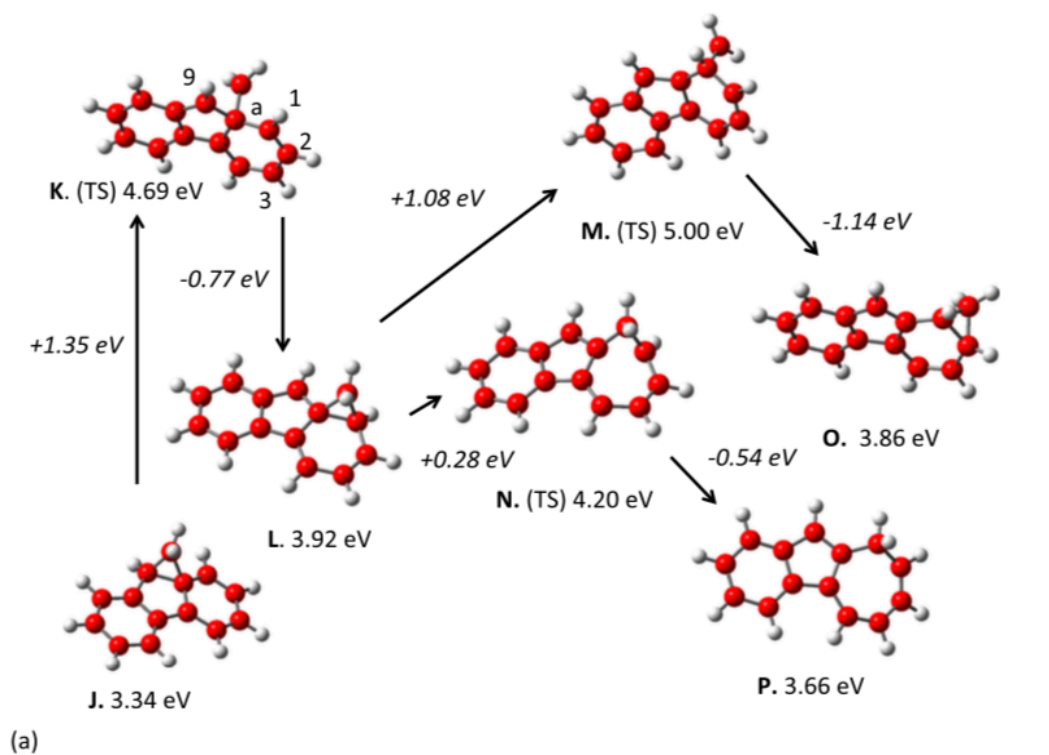

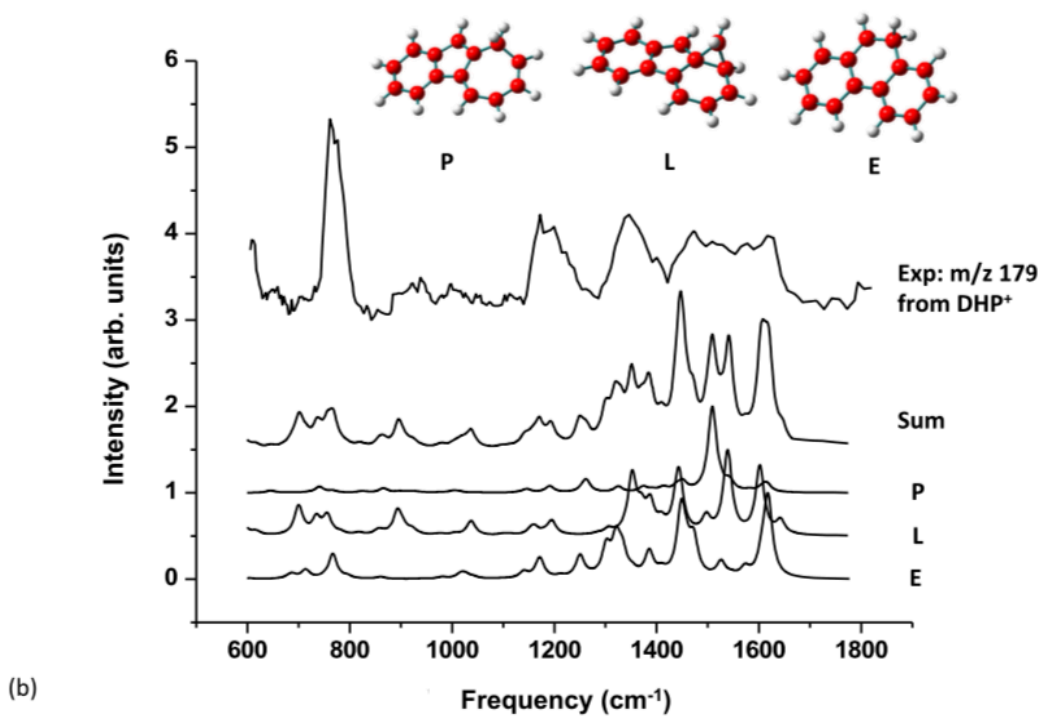

**Figure 5.** (a) Potential energy diagram for the isomerization of 10-hydrophenanthrene, E. Either pathway (in Figure 4(b)) produces the 9-methylenefluorene ion, I, which is most likely in equilibrium with J, in this figure. Here the methylene group has formed a three-membered ring attached to the five-membered ring. Proceeding through TS K involves a shift of the methylene group to form L, the (a,1)-cyclomethylenefluorenyl ion. Two

routes are now possible: a further shift of methylene proceeding through TS M to O, the (1,2)-cyclomethylenefluorenyl ion, or opening the three-membered ring in L and producing the 1-hydrocycloheptatriene-2,3-indenyl ion P via the TS structure N. Though not pictured, the three-membered ring in O can open to form the 2-hydrocycloheptatriene-2,3-indenyl ion. (b) Experimental infrared multiple-photon dissociation (IRMPD) spectra of the *m/z* = 179 fragment ion from DHP$^+$ (top) compared to the calculated spectra of the 10-hydrophenanthrene ion [DHP–H(9)]$^+$ E, (a,1)-cyclomethylenefluorenyl ion L, and 1-hydrocycloheptatriene-2,3-indenyl ion P, and their sum. All computed spectra (B3LYP/6-31G (d, p)) were scaled by 0.97 and a spectral bandwidth of 10 cm$^{-1}$ was adopted.